\documentstyle[12pt]{article}

\begin{document}
\begin{center}
{\large PONTRYAGIN AND EULER FORMS AND CHERN-SIMONS TERMS
        IN WEYL-CARTAN SPACE}\\
\vskip 0.4cm
O. V. Babourova and B. N. Frolov\\
{\it Department of Mathematics, Moscow State Pedagogical University,\\
     Krasnoprudnaya 14, Moscow 107140, Russia;\\
     E-mail: baburova.physics@mpgu.msk.su,\enspace frolovbn.physics@mpgu.msk.su}
\vskip 1cm
\end{center}

\begin{abstract}
{\small
\par
        The existence of the Pontryagin and Euler forms in a Weyl-Cartan
space on the basis of the variational method with Lagrange multipliers are
established. It is proved that these forms can be expressed via the exterior
derivatives of the corresponding Chern-Simons terms in a Weyl-Cartan space
with torsion and nonmetricity.
}
\end{abstract}

PACS number(s): 04.20.Fy, 04.50.+h
\newpage
\renewcommand{\thesection}{\Roman{section}.}
\section {Introduction}
\renewcommand{\thesection}{\arabic{section}}
\markright{Pontryagin and Euler forms and Chern-Simons...}
\par
        The dilatonic gravity is one of the attractive approaches to the
modern gravitational theory. This is based on the fact that a low-energy
effective string theory is reduced to the theory of interacting metric and
scalar dilaton field.$^{1}$ From the geometric point of view the
dilatonic gravity is connected with the Weyl geometry of space-time.
\par
        The attractive feature of any quantum field theory is its
renormalizability. It is well known that the Einstein gravity coupled to a
scalar field is nonrenormalizable at the one-loop level.$^{2}$ In
order to get renormalizability one can add to the gravitational Lagrangian
curvature-squared terms.$^{3}$ In this connection the Gauss-Bonnet type
identity for quadratic Lagrangians becomes the object of a considerable
amount of attention. The generalization of the Gauss-Bonnet formula to
four-dimensional psoudo-riemannian space $V_{4}$ was performed by Bach$^{4}$
and Lanczos$^{5}$ and on the basis of the variational method
with Lagrange multipliers by Ray.$^{6}$ It is well known that
Bach-Lanczos identity in a Riemann space implies the one-loop
renormalizability of pure gravitation.$^{2}$ The generalization of the
Bach-Lanczos identity to a Riemann-Cartan space $U_{4}$ was performed
in Refs. 7-9.
\par
        The Gauss-Bonnet-Chern Teorem$^{10}$ states that an integral
of some form (called Euler form) over an oriented compact manifold without
boundary does not depend on the choice of a metric and a connection of the
manifold and therefore is the topological invariant of the manifold (its
Euler characteristic). There exists another class of topological invariants
of a manifold represented as integrals of Chern-Pontryagin forms over the
manifold.$^{11}$ These topological invariants are closely connected with
the so-called topological charges of the space-time manifolds.
\par
        We shall obtain the Pontryagin type and the Euler type topological
invariants in a Weyl-Cartan space $Y_{4}$ (the generalized Weyl space with
torsion), which will be essential for the dilatonic gravitational theory
with quadratic Lagrangians in the spaces with torsion and nonmetricity.

\renewcommand{\thesection}{\Roman{section}.}
\section{Topological invariants in general metric-affine space}
\renewcommand{\thesection}{\arabic{section}}
\markright{Pontryagin and Euler forms and Chern-Simons...}
\setcounter{equation}{0}
        In this section we consider a metric-affine space ($L_{4}$,$g$) that
is a connected 4-dimensional oriented differentiable manifold ${\cal M}$
equipped with a linear connection $\Gamma$ and a metric $g$ of index 1.$^{12}$
We shall use an anholonomic local vector frame $e_{a}$
($a=1,2,3,4$) and a 1-form coframe $\theta^{a}$ with $e_{a}\rfloor\theta^{b}
=\delta^{b}_{a}$ ($\rfloor$ means the interior product). The vector basis
$e_{a}$ can be chosen to be pseudo-orthonormal with respect to a metric
\begin{equation}
g=g_{ab}\theta^{a}\otimes\theta^{b}\;.\label{eq:0}
\end{equation}
In this case one gets,
\begin{equation}
g_{ab} := g(e_{a},e_{b}) = diag(+1,+1,+1,-1)\;. \label{eq:1}
\end{equation}
\par
        In ($L_{4}$,$g$) a metric $g$ and a connection $\Gamma$ are not
compatible in the sense that the $GL(4,R)$-covariant exterior differential
(${\cal D}:=d + \Gamma\wedge\ldots$) of the metric does not vanish,
\begin{equation}
{\cal D}g_{ab} = dg_{ab} - \Gamma^{c}\!_{a}g_{cb} - \Gamma^{c}\!_{b}g_{ac}
=: - Q_{ab}\; , \label{eq:2}
\end{equation}
where $\Gamma^{a}\!_{b}$ is a connection 1-form and $Q_{ab}$ is a
nonmetricity 1-form.
\par
        A curvature 2-form $\Omega^{a}\!_{b}$ and a torsion 2-form
${\cal T}^{a}$,
\begin{equation}
\Omega^{a}\!_{b}=\frac{1}{2}R^{a}\!_{bcd}\theta^{c}\wedge\theta^{d}\;,
\qquad {\cal T}^{a}=\frac{1}{2}T^{a}\!_{bc}\theta^{b}\wedge\theta^{c}\;,
\label{eq:3}
\end{equation}
are defined by virtue of the Cartan's structure equations,
\begin{eqnarray}
\Omega^{a}\!_{b}=d\Gamma^{a}\!_{b}+\Gamma^{a}\!_{c}\wedge\Gamma^{c}\!_{b}\;,
\label{eq:4}\\ {\cal T}^{a}={\cal D}\theta^{a}=d\theta^{a}+\Gamma^{a}\!_{b}
\wedge\theta^{b}\;. \label{eq:5}
\end{eqnarray}
\par
        Let us consider the 4-form
\begin{equation}
\Pi= B^{b}\!_{a}\!^{q}\!_{p}\Omega^{a}\!_{b}\wedge\Omega^{p}\!_{q}
\;,  \label{eq:6}
\end{equation}
where $B^{b}\!_{a}\!^{q}\!_{p}$ is an unknown $GL(4,R)$-invariant tensor.
From the form of (\ref{eq:6}) it is easy to get the following symmetry
property of this tensor,
\begin{equation}
B^{b}\!_{a}\!^{q}\!_{p}=B^{q}\!_{p}\!^{b}\!_{a}\;. \label{eq:06}
\end{equation}
The 4-form (\ref{eq:6}) is proportional to the volume 4-form $\eta$ of the
4-dimensional manifold ${\cal M}$, where
\begin{equation}
\eta = \frac{1}{4!}\eta_{abcd}\theta^{a}\wedge\theta^{b}\wedge \theta^{c}
\wedge\theta^{d}\;, \quad \eta_{abcd}=\sqrt{-det \Vert g_{kl}\Vert}\,
\epsilon_{abcd}\;.
\end{equation}
Here $\epsilon_{abcd}$ is the components of the totaly antisymmetric
$GL(4,R)$-invariant Levi-Civita 4-form density $^{12}$
($\epsilon_{1234}=-1$).
\par
        Since ${\cal D}\eta = d\eta = 0$ as a 5-form on the 4-dimensional
manifold ${\cal M}$, one has the identity$^{13}$
\begin{equation}
{\cal D}\eta_{abcd} = -\frac{1}{2} Q \eta_{abcd}\;, \qquad Q:=g^{pq} Q_{pq}
\;. \label{eq:290}
\end{equation}
\par
        The explicit form of the tensor (\ref{eq:06}) should be determined on
the basis of the condition that the integral
\begin{equation}
\int_{{\cal M}}\Pi \label{eq:7}
\end{equation}
over the oriented 4-demensional manifold {\cal M} without boundary
does not depend on the choise of a metric and a connection and therefore
the variation of the integrand of (\ref{eq:7}) with respect to a metric and a
connection should be equal to an exact form. Here we consider the manifold
${\cal M}$ without boundary for the simplicity. For the manifolds with
boundary some additional surface terms should be taken into account.$^{11}$
\par
        As a consequence of (\ref{eq:0}) the variation with respect to a
metric $g$ is determined only by variations of 1-forms $\theta_{a}$ because
of the fact that the variation $\delta g_{ab}=0$ when one chooses the
pseudo-orthonormal basis $e_{a}$ and gets the condition (\ref{eq:1}). The
tensor $B^{b}\!_{a}\!^{q}\!_{p}$ in (\ref{eq:6}) also should not to be varied
when the local vector basis $e_{a}$ is chosen to be anholonomic and
pseudo-orthonormal because it can be constructed (as an $GL(4,R)$-invariant
tensor) only from the metric tensor, Kronecker delta $\delta^{b}_{a}$ and the
$GL(4,R)$-invariant totaly antisymmetric Levi-Civita density
$\epsilon_{abpq}$.
\par
        The variation of (\ref{eq:6}) yields the expression,
\begin{equation}
\delta\Pi=2\delta\Gamma^{a}\!_{b}\wedge ({\cal D} B^{b}\!_{a}\!^{q}\!_{p})
\wedge\Omega^{p}\!_{q}+d(2\delta\Gamma^{a}\!_{b}\wedge
 B^{b}\!_{a}\!^{q}\!_{p} \Omega^{p}\!_{q})\;.\label{eq:8}
\end{equation}
Here the following relation has been used,
\begin{equation}
\delta\Omega^{a}\!_{b}\wedge\Phi^{b}\!_{a}=d(\delta\Gamma^{a}\!_{b}\wedge
\Phi^{b}\!_{a}) + \delta\Gamma^{a}\!_{b}\wedge {\cal D}\Phi^{b}\!_{a}\; ,
\label{eq:9}
\end{equation}
that valids for an arbitrary 2-form $\Phi^{a}\!{_b}$.
\par
        One can see that the variation (\ref{eq:8}) is equal to an exact
form, if the tensor $B^{b}\!_{a}\!^{q}\!_{p}$ satisfies the condition,
\begin{equation}
{\cal D} B^{b}\!_{a}\!^{q}\!_{p}= 0 \;. \label{eq:11}
\end{equation}
\par
In a general metric-affine space ($L_{4}$,$g$) there are only two
possibilities to satisfy (up to constant factors) the condition (\ref{eq:11}),
\begin{equation}
(a)\quad B^{b}\!_{a}\!^{q}\!_{p} =\delta^{b}_{a}\delta^{q}_{p}\;, \qquad
(b)\quad B^{b}\!_{a}\!^{q}\!_{p} =\delta^{b}_{p}\delta^{q}_{a}\;.\label{eq:12}
\end{equation}
\par
        In the case (a) the 4-form $\Pi$ (\ref{eq:6}) reads,
\begin{equation}
\Pi_{\Omega}=\Omega^{a}\!_{b}\wedge\Omega^{b}\!_{a} = \mbox{\rm Tr}(\Omega
\wedge\Omega )\;, \label{eq:13}
\end{equation}
and in the case (b) one has,
\begin{equation}
\Pi_{tr\Omega}=\Omega^{a}\!_{a}\wedge \Omega^{b}\!_{b}= \mbox{\rm Tr}\Omega
\wedge\mbox{\rm Tr}\Omega \;. \label{eq:15}
\end{equation}
We see that the 4-forms (\ref{eq:13}) and (\ref{eq:15}) are equal up to
constant factors to the well-known Pontryagin forms.$^{11,12}$

\renewcommand{\thesection}{\Roman{section}.}
\section{The topological invariants in a Weyl- \newline Cartan space}
\renewcommand{\thesection}{\arabic{section}}
\markright{Pontryagin and Euler forms and Chern-Simons...}
\setcounter{equation}{0}
        A Weyl-Cartan space $Y_{4}$ is a space with a metric,
curvature, torsion and nonmetricity which obeys the constraint,
\begin{equation}
Q_{ab} = \frac{1}{4}g_{ab}Q\;.  \label{eq:21}
\end{equation}
This constraint can be introduced into the variational approach with the help
of the method of Lagrange multipliers. In this case the integral (\ref{eq:7})
has to be modified,
\begin{equation}
\int_{{\cal M}}\left (\Pi + \Lambda^{ab}\wedge (Q_{ab} - \frac{1}{4}g_{ab}Q)
\right )\;, \label{eq:22}
\end{equation}
where the Lagrange multiplier $\Lambda^{ab}$ is a tensor-valued
3-form with the properties,
\begin{equation}
\Lambda^{ab}=\Lambda^{ba}\;, \qquad   \Lambda^{a}\!_{a}=0 \;. \label{eq:220}
\end{equation}
\par
        The variation of (\ref{eq:22}) with respect to $\theta^{a}$,
$\Gamma^{a}\!_{b}$ and the Lagrange mutiplier yields that the following
variational derivatives have to vanish identically,
\begin{eqnarray}
\delta \Gamma^{a}\!_{b}\;: &{\cal D}(B^{b}\!_{a}\!^{q}\!_{p})\wedge
\Omega^{p}\!_{q} - \Lambda^{b}\!_{a} = 0 \;, & \label{eq:23}\\
\delta\Lambda^{ab}\;: &Q_{ab} - \frac{1}{4}Q g_{ab} = 0 \;.
&\label{eq:24}
\end{eqnarray}
As in the previous section the variational derivative with respect to
$\theta^{a}$ is absent because of the fact that there is no an explicit
dependence on $\theta^{a}$ of the integrand expression in (\ref{eq:22}).
\par
        The identity (\ref{eq:23}) in $Y_{4}$ is equivalent to the following
identities,
\begin{eqnarray}
&\Lambda^{ba}= ({\cal D} - \frac{1}{4}Q)B^{(ba)q}\!_{p} \wedge
\Omega^{p}\!_{q}\; , \label{eq:25}\\
&({\cal D} - \frac{1}{4}Q)B^{[ba]q}\!_{p} \wedge
\Omega^{p}\!_{q}= 0\; , \label{eq:26}\\
&{\cal D}B^{a}\!_{a}\!^{q}\!_{p} \wedge\Omega^{p}\!_{q} = 0\;. \label{eq:27}
\end{eqnarray}
\par
        The identities (\ref{eq:26}), (\ref{eq:27}) are satisfied in
the following four cases:
\begin{eqnarray}
&(a)\quad B^{baq}\!_{a}\!^{q}\!_{p} =g^{ba}\delta^{q}_{p}\;, \qquad
(b)\quad B^{baq}\!_{p} =\delta^{b}_{p}g^{qa}\;,
\label{eq:28}\\
&(c)\quad B^{baq}\!_{p} =g^{bq}\delta^{a}_{p}\;, \qquad
(d)\quad B^{baq}\!_{p} =\eta^{baq}\!_{p}\;.\label{eq:29}
\end{eqnarray}
In the case (d) one has to use (\ref{eq:2}), (\ref{eq:21}) and
(\ref{eq:290}).
\par
        The equality (\ref{eq:25}) determines the Lagrange miltiplier.
In all cases (a)-(d) one has $\Lambda^{ab}=0$. This means that the
Weyl-Cartan constraint (\ref{eq:24}) can be imposed both before and after the
variational procedure.
\par
        The cases (a) and (b) coinside with (\ref{eq:12}) and yield for a
Weyl-Cartan space $Y_{4}$ the Pontryagin forms (\ref{eq:13}) and (\ref{eq:15})
of the previous section. The cases (c) and (d) appear in $Y_{4}$ but not in
($L_{4}$,$g$).
\par
        In the case (c) one has the Pontryagin form,
\begin{equation}
\Pi_{CW}=\Omega^{ab}\wedge \Omega_{ab} = \mbox{\rm Tr}(\Omega \wedge \Omega
^{T})\;,\label{eq:30}
\end{equation}
where $\Omega^{T}$ means the transpose of $\Omega$. In $Y_{4}$ with the
help of the relation,
\begin{equation}
\Omega_{ab}= \Omega_{[ab]} + \frac{1}{4}g_{ab}\mbox{\rm Tr}\Omega\;,
\label{eq:31}
\end{equation}
(\ref{eq:30}) can be decomposed as follows,
\begin{equation}
\Omega^{ab}\wedge \Omega_{ab}=\Omega^{[ab]}\wedge \Omega_{[ab]} + \frac{1}{4}
\mbox{\rm Tr}\Omega \wedge \mbox{\rm Tr}\Omega\;. \label{eq:32}
\end{equation}
On the other hand the Pontryagin form (\ref{eq:13}) in $Y_{4}$ has the
decomposition,
\begin{equation}
\Omega^{a}\!_{b}\wedge \Omega^{b}\!_{a}= -\Omega^{[ab]}\wedge\Omega_{[ab]}
+\frac{1}{4}\mbox{\rm Tr}\Omega\wedge \mbox{\rm Tr}\Omega \;. \label{eq:33}
\end{equation}
Therefore in a Weyl-Cartan space $Y_{4}$ one has two fundamental Pontryagin
forms, which are equal up to constant factors to
\begin{equation}
\Pi_{C}=\Omega^{[ab]}\wedge \Omega_{[ab]}\;, \qquad
\Pi_{W}=\mbox{\rm Tr}\Omega\wedge \mbox{\rm Tr}\Omega \;.
\label{eq:34}
\end{equation}
The former form is the volume preserving Pontryagin form and the latter one
is the dilatonic Pontryagin form.
\par
    In the case (d) we get the Euler form in a Weyl-Cartan space $Y_{4}$,
\begin{equation}
{\cal E}= \eta^{b}\!_{a}\!^{q}\!_{p}\Omega^{a}\!_{b}\wedge \Omega^{p}\!_{q}
\;. \label{eq:35}
\end{equation}
One can use the holonomic coordinate basis $e_{\alpha} = \partial_{\alpha}$
and express the topological invariant corresponding to (\ref{eq:35}) in the
component form,
\begin{equation}
\int_{{\cal M}}{\cal E} = \int_{{\cal M}}E \sqrt{-g}dx^{1}\wedge dx^{2}
\wedge dx^{3}\wedge dx^{4}\;, \label{eq:36}
\end{equation}
\begin{equation}
E =  R^{2} - (R_{\alpha\beta} +\tilde{R}_{\alpha\beta})(R^{\beta\alpha} +
\tilde{R}^{\beta\alpha}) + R_{\alpha\beta\mu\nu}R^{\mu\nu\alpha\beta}\; ,
\label{eq:37}
\end{equation}
where $R^{\alpha}\!_{\beta\mu\nu}$ are the components of the curvature 2-form
in a holonomic basis, the following notations being used,
$R_{\alpha\beta} = R^{\sigma}\!_{\alpha\sigma\beta}$,
$\tilde{R}_{\alpha\beta} = R_{\alpha\sigma\beta}\!^{\sigma}$,
$R = R_{\sigma}\!^{\sigma}$.
\par
        The Gauss-Bonnet-Chern Teorem$^{10}$ states the relation of the
integral (\ref{eq:36}) over the oriented compact manifold ${\cal M}$ without
boundary with the Euler characteristic of this manifold.
The explicit proof using a holonomic basis of the independence of
(\ref{eq:36}) on the choice of a metric and a connection of
a Weyl-Cartan space $Y_{4}$ is explained in Ref. 14.

\renewcommand{\thesection}{\Roman{section}.}
\section{Chern-Simons terms in a Weyl- \newline Cartan space}
\renewcommand{\thesection}{\arabic{section}}
\markright{Pontryagin and Euler forms and Chern-Simons...}
\setcounter{equation}{0}
        It is well known that in ($L_{4}$,$g$) the Pontryagin forms  can be
represented as the exterior derivatives of the $GL(4,R)$ Chern-Simons
terms,$^{12}$
\begin{eqnarray}
&\Pi_{\Omega}=d{\cal C}_{\Omega}\;, \qquad {\cal C}_{\Omega}=\Gamma^{b}\!_{a}
\wedge \Omega^{a}\!_{b} - \frac{1}{3}\Gamma^{b}\!_{a}\wedge\Gamma^{a}\!_{c}
\wedge \Gamma^{c}\!_{b}\;, \label{eq:40}\\
&\Pi_{W}=d{\cal C}_{W}\;, \qquad {\cal C}_{W} = \frac{1}{2}Q\wedge \Omega^{a}
\!_{a}\;. \label{eq:41}
\end{eqnarray}
\par
        It is easy to see that Pontryagin form $\Pi_{C}$ (\ref{eq:34}) in a
Weyl-Cartan space $Y_{4}$ can be represented in an analogous manner,
\begin{equation}
\Pi_{C} = d{\cal C}_{C}\;, \qquad {\cal C}_{C}=
\Gamma^{[b}\!_{a]}\wedge \Omega^{[a}\!_{b]} - \frac{1}{3}\Gamma^{[b}\!_{a]}
\wedge\Gamma^{[a}\!_{c]}\wedge \Gamma^{[c}\!_{b]}\;. \label{eq:42}
\end{equation}
\par
        As it was pointed out in Ref. 12, the Euler form (\ref{eq:35})
in the framework of a Riemann-Cartan space can be expressed in terms of the
corresponding Chern-Simons type construction,
\begin{equation}
{\cal E} = d{\cal C}_{{\cal E}}\;, \qquad {\cal C}_{{\cal E}} =
\eta^{b}\!_{a}\!^{q}\!_{p} \left (\Omega^{a}\!_{b}\wedge \Gamma^{p}\!_{q} -
\frac{1}{3}\Gamma^{a}\!_{b}\wedge \Gamma^{p}\!_{f}\wedge\Gamma^{f}\!_{q}
\right ) \;. \label{eq:43}
\end{equation}
\par
        Let us prove that formula (\ref {eq:43}) is also valid in a
Weyl-Cartan space $Y_{4}$. The proof is based on the two Lemmas.
\par
        {\it Lemma 1}. If the equality
\begin{equation}
{\cal D}\eta^{b}\!_{a}\!^{q}\!_{p}=0\;, \label{eq:44}
\end{equation}
is valid, then the identity (\ref{eq:43}) is fulfilled.
\par
        {\it Proof}. In anholonomic orthonormal frames one has
$d\eta^{b}\!_{a}\!^{q}\!_{p}=0$, and therefore (\ref{eq:44}) yields,
\begin{equation}
\Gamma^{b}\!_{f}\eta^{f}\!_{a}\!^{q}\!_{p}-\Gamma^{f}\!_{a}\eta^{b}\!_{f}\!
^{q}\!_{p}+\Gamma^{q}\!_{f}\eta^{b}\!_{a}\!^{f}\!_{p}-\Gamma^{f}\!_{p}
\eta^{b}\!_{a}\!^{q}\!_{f}=0\;. \label{eq:45}
\end{equation}
After multiplying (\ref{eq:45}) externally by the 3-form $\Gamma^{a}\!_{s}
\wedge\Gamma^{s}\!_{b}\wedge\Gamma^{p}\!_{q}\wedge$, one gets the
$Y_{4}$-identity,
\begin{equation}
\eta^{b}\!_{a}\!^{q}\!_{p}\Gamma^{a}\!_{s}\wedge\Gamma^{s}\!_{b}\wedge
\Gamma^{p}\!_{f}\wedge\Gamma^{f}\!_{q}=0\;. \label{eq:46}
\end{equation}
After multiplying (\ref{eq:45}) externally by the 3-form $\Omega^{a}\!_{b}
\wedge\Gamma^{p}\!_{q}\wedge$, one gets the second $Y_{4}$-identity,
\begin{equation}
\eta^{b}\!_{a}\!^{q}\!_{p}(2\Omega^{a}\!_{b}\wedge\Gamma^{p}\!_{f}\wedge
\Gamma^{f}\!_{q}-\Omega^{a}\!_{f}\wedge\Gamma^{f}\!_{b}\wedge\Gamma^{p}\!_{q}
+\Gamma^{a}\!_{f}\wedge\Omega^{f}\!_{b}\wedge\Gamma^{p}\!_{q})=0\;.
\label{eq:47}
\end{equation}
Now using the identities (\ref{eq:46}) and (\ref{eq:47}), the Cartan's
structure equation (\ref{eq:4}) and the Bianchi identity,
\begin{equation}
{\cal D}\Omega^{a}\!_{b}=d\Omega^{a}\!_{b}+\Gamma^{a}\!_{f}\wedge
\Omega^{f}\!_{b}-\Omega^{a}\!_{f}\wedge\Gamma^{f}\!_{b}=0\;,
\end{equation}
let us perform the exterior differentiation of the Chern-Simons term
${\cal C}_{{\cal E}}$ (\ref{eq:43}) and get,
\begin{eqnarray}
&d{\cal C}_{{\cal E}}-{\cal E}=\frac{1}{3}\eta^{b}\!_{a}\!^{q}\!_{p}
\Gamma^{a}\!_{s}\wedge\Gamma^{s}\!_{b}\wedge\Gamma^{p}\!_{f}\wedge
\Gamma^{f}\!_{q}\nonumber \\
&-\frac{2}{3}\eta^{b}\!_{a}\!^{q}\!_{p}
(2\Omega^{a}\!_{b}\wedge\Gamma^{p}\!_{f}\wedge\Gamma^{f}\!_{q}
-\Omega^{a}\!_{f}\wedge\Gamma^{f}\!_{b}\wedge\Gamma^{p}\!_{q}
+\Gamma^{a}\!_{f}\wedge\Omega^{f}\!_{b}\wedge\Gamma^{p}\!_{q})=0\;,
\end{eqnarray}
as was to be proved.
\par
        {\it Lemma 2}. The equality (\ref{eq:44}) is valid if and only if
the space under consideration is a Weyl-Cartan space $Y_{4}$.
\par
        {\it Proof}. In a general ($L_{4}$,$g$) space one has,
\begin{equation}
{\cal D}\eta^{b}\!_{a}\!^{q}\!_{p}=\eta^{q}\!_{map}\tilde{Q}^{bm} -
\eta^{b}\!_{map}\tilde{Q}^{qm}\;, \label{eq:48}
\end{equation}
where $\tilde{Q}^{bm} := Q^{bm}- \frac{1}{4}g^{bm}{Q}$ is
the tracefree part of the nonmetricity 1-form, $\tilde{Q}^{b}\!_{b}
=0$. For a Weyl-Cartan space $Y_{4}$ one has $\tilde{Q}^{bm}=0$ and
the sufficient condition of the Lemma is evident. The necessary condition of
the Lemma is the consequence of the fact that the vanishing of (\ref{eq:48})
leads to the equality,
\begin{equation}
g^{ab}\tilde{Q}^{pq}-g^{bp}\tilde{Q}^{aq}-g^{aq}
\tilde{Q}^{bp}+g^{pq}\tilde{Q}^{ab}=0\;,
\end{equation}
which yields $\tilde{Q}^{pq}=0$, as was to be proved.

\renewcommand{\thesection}{\Roman{section}.}
\section{Conclusions}
\renewcommand{\thesection}{\arabic{section}}
\markright{Pontryagin and Euler forms and Chern-Simons...}
        We have proved the existence of the Pontryagin and Euler forms in a
Weyl-Cartan space on the basis of the variational method with Lagrange
multipliers. It has been discovered that the  Pontryagin form,
$\Pi_{C}=\Omega^{[ab]}\wedge \Omega_{[ab]}$, and Euler form,
${\cal E}= \eta^{b}\!_{a}\!^{q}\!_{p}\Omega^{a}\!_{b}\wedge \Omega^{p}\!_{q}$,
which are specific for a Riemann-Cartan space, also exist in a Weyl-Cartan
space. With the help of these forms the topological invariants of a
Weyl-Cartan space which do not depend on the choice of a metric and a
connection are constructed. It has been proved that these forms can be
expressed via the exterior derivatives of the corresponding Chern-Simons
terms in a Weyl-Cartan space (see (\ref{eq:42}) and (\ref{eq:43}),
respectively). From the Lemma 2 proved it follows that the relation
(\ref{eq:43}) is not valid in the more general geometry than the Weyl-Cartan
one.
\newpage
\vskip 0.6cm
\begin{description}
\item{$^{1}$}
M.B. Green,J.H. Schwarz and E. Witten, {\em Superstring Theory}, 2 volumes
(Cambridge University Press, Cambridge, 1987).
\item{$^{2}$}
G. 't Hooft, M. Veltman, {\em Ann. Inst. H. Poincar\'{e}} {\bf 20}, 69 (1974).
\item{$^{3}$}
K.S. Stelle, {\em Phys. Rev.} {\bf D16}, 953 (1977).
\item{$^{4}$}
R. Bach, {\em Math. Z.} {\bf 9}, 110 (1921).
\item{$^{5}$}
C. Lanczos, {\em Ann. Math.(N.Y.)} {\bf 39}, 842 (1938).
\item{$^{6}$}
J.R. Ray, {\em J. Math. Phys.} {\bf 19}, 100 (1978).
\item{$^{7}$}
V.N. Tunjak, {\em Izvestija vyssh. uch. zaved. (Fizika)} N9, 74 (1979) [in
Russian].
\item{$^{8}$}
H.T. Nieh, {\em J. Math. Phys.} {\bf 21}, 1439 (1980).
\item{$^{9}$}
K. Hayashi, T. Shirafuji, {\em Prog. Theor. Phys.} {\bf 65}, 525 (1981).
\item{$^{10}$}
R. Sulanke und P. Wintgen, {\em Differentialgeometrie und faserb\"undel}
(Hoch\-schulb\"ucher f\"ur mathematik, band 75)(VEB Deutscher Verlag der
Wis\-senshaften, Berlin, 1972).
\item{$^{11}$}
T. Eguchi, P.B. Gilkey and A.J. Hanson, {\em Phys. Reports} {\bf 66}, 213
(1980).
\item{$^{12}$}
F.W. Hehl, J.D. McCrea, E.W. Mielke and Yu. Ne'eman, {\em Phys. Reports}
{\bf 258}, 1 (1995).
\item{$^{13}$}
R. Tresguerres, {\em J. Math. Phys.} {\bf 33}, 4231 (1992).
\item{$^{14}$}
O.V. Babourova, B.N. Frolov, {\em Gauss-Bonnet type identity in Weyl-Cartan
space-time}, LANL e-archive gr-qc/9608... (1996).
\end{description}
\end{document}